
\def\beginsection#1{\vskip .25in\goodbreak\noindent
		    {\bf{#1}}\nobreak
		    \smallskip\nobreak\noindent\hskip -.0mm}


\def\ga{\gamma}
\def\de{\delta}         
\def\ep{\epsilon}

\def\la{\lambda}        \def\La{\Lambda}

         \def\Si{\Sigma}

         \def\Om{\Omega}

\def\F{\hat F}
\def\G{\nabla\nu}
\def\T{\hat\nabla T}
\def\W{\hat\nabla W}

\def\pmb#1{\setbox0=\hbox{#1}%
    \kern-.025em\copy0\kern-\wd0\kern.05em\copy0\kern-\wd0
    \kern-.025em\raise.0433em\box0 }
\def\g{\pmb{\hbox{{\it g}}}}
\def\LG{$\widehat{LG}$}
\def\lg{\widehat{l\g}}
\def\tr{{\rm tr}}
\def\gh{\hat g}
\def\Fh{\hat F}

\def\Ah{\hat A}
\def\Ax{A_x}
\def\Abar{\bar A}
\def\d#1{\,\partial_{{#1}}}
\def\dbar{\,\bar\partial}
\def\Dbar{\bar D}

\def\eom{equations of motion}
\def\wk{Weyl-Kac}
\def\vir{Virasoro}
\def\e#1{e^{{#1}}}
\def\Pe#1{{\rm P}e^{{#1}}}
\def\bra{\langle}
\def\ket{\rangle}
\def\csa{Cartan subalgebra}

\def\reals{{\it I\!\!R}}

\def\mc{g^{-1}dg}
\def\half{\textstyle {1\over2}}

\def\rep{representation}
\def\chr{character}

\def\intxt{\int dx\,dt\,}
\def\ad{{\rm ad\,}}

\def\t#1{\dot{#1}}
\def\ts{\textstyle}
\def\4{{\ts{1\over4\pi}}\tr\int}
\def\four{{\ts{i\over4\pi}}\tr\int}
\def\fourk{{\ts{ik\over4\pi}}\tr\int}
\def\2{{\ts{1\over2\pi}}\tr\int}
\def\twelve{{\ts{ik\over12\pi}}\tr\int}
\def\two{{\ts{i\over2\pi}}\tr\int}
\def\twok{{\ts{ik\over2\pi}}\tr\int}
\def\ct{{\ts{\hbox{{\it c}}\over24\pi}}}
\def\tc{{\ts{24\pi\over c}}}
\def\cf{{\ts{\hbox{{\it c}}\over48\pi}}}
\def\diffs{{\rm diff}S^1}
\def\xt{dx\,dt\,}
\def\intsx{\int_\Si\int dx\,}
\font\sm=cmsy7
\def\c{{\sm\circ}}
\def\L_0{{\cal L}_0}
\def\w{$W$-algebra}
\def\ac{affine connection}
\def\km{Kac-Moody}
\def\wzw{WZW model}
\def\cs{Chern-Simons theory}
\def\pw{Polyakov-Wiegmann}
\def\3{three dimension}

\def\wi{Ward identity}
\def\cft{conformal field theory}
\def\note{paper}
\def\tft{topological field theory}
\def\pin{path integral}
\def\wl{Wilson loop}
\def\gd{Gelfand-Dickey}
\def\twod{two dimension}
\def\v{\hbox{{\sl Vir}}}

\vskip 1in
\centerline{\bf THREE DIMENSIONAL FIELD THEORIES FROM}
\centerline{\bf INFINITE DIMENSIONAL LIE ALGEBRAS}

\vskip 1.5in
\centerline{R.E.C. Perret}
\smallskip
\count0=0

\vskip .5cm
\centerline{\it Department of Physics}
\centerline{\it University of Southern California}
\centerline{\it Los Angeles, CA 90089--0484}
\vskip 1in
\centerline{Submitted to: \it Nuclear Physics B}

\vskip .5in

\beginsection{Abstract:}
A procedure for constructing topological actions
from centrally extended Lie groups
is introduced.
For a \km\ group, this produces \3al \cs,
while for the \vir\ group the result is a new \3al \tft\
whose physical states satisfy the \vir\ \wi.
This \tft\ is shown to be a first order formulation of two
dimensional induced gravity in the chiral gauge.
The extension to $W_3$-gravity is discussed.

\vskip .5in\noindent
USC--92/018\hfill September 1992

\vskip.5in\eject

\def\Witi{1}
\def\Witii{2}
\def\BN{3}
\def\EV{4}
\def\FAS{5}
\def\Stone{6}
\def\AS{7}
\def\Wieg{8}
\def\WK{9}
\def\walg{10}
\def\V{11}
\def\Witiii{12}
\def\IZ{13}
\def\P{14}
\def\SvN{15}

\beginsection{1. Introduction}
In recent years, numerous connections have been discovered between
conformal field theories, integrable models,
quantum groups and knot theory.
A particularly illuminating framework for discussing
these  relationships is \cs\ in \3s [\Witi].
 \cs\ provides an intrinsically \3al
description of many knot invariants and elucidates
the relationship between knot
theory and two dimensional statistical mechanics [\Witii].

The starting point of this development
was the realization by Witten that there is an intimate connection
between \3al \cs\ and two dimensional current algebra [\Witi].
The space of conformal blocks of the \wzw\ coincides with the
space of physical states in the canonical quantization of \cs\ [\Witi,\BN].
This \3al interpretation of the conformal blocks makes manifest
some surprising symmetry
of their braiding matrices [\Witii].
It also gives, in the form of Wilson lines,
an explicit representation of the Verlinde operators [\EV],
which play an important role in the study of the modular properties
of the conformal blocks [\Witi].

In this \note\ I will show that the relationship between \cs\ and
the \wzw\ can be understood directly from the underlying chiral algebra,
the affine \km\ algebra for a compact Lie group $G$. As is well known,
one can express a character of a Lie algebra as a coherent state \pin\
[\FAS,\Stone,\AS].
The action in this \pin\ is the geometric action on the coadjoint orbit
of the corresponding group---in the case of an affine \km\ algebra,
the action of the chiral \wzw\ [\AS,\Wieg,\WK].
This action does not define a consistent
quantum field theory in its own right;
rather, the corresponding (formally defined)
partition function coincides with the
chiral building blocks of the partition function of the full \cft\ [\WK].

In Section~2, I
formally consider the trace of the path ordered
exponential of a two dimensional gauge field for a \km\ group.
Constructing a geometric action
in exactly the same way as in the case of a character,
I arrive at a gauged
chiral \wzw\ coupled to a \3al \cs. The third component of the
\3al gauge field is associated with the central extension of the algebra.
The symmetry between all three components is
the main nontrivial feature of the construction, and points to
some hidden symmetry of the \km\ algebra.

It is an interesting question to what extent this construction
can be applied to other chiral algebras.
If the chiral algebra
of a \cft\ can be related to a \tft\ in which Wilson like
observables  are defined, one would expect the latter to be
related to the conformal blocks. Also,
the expectation values of such observables
might give rise to a nontrivial
knot invariant.

In Section~3, I carry out the analogous construction for
the \vir\ algebra.
As in the \km\ case there is indeed a nontrivial
\3al symmetry, which is discussed in Section~4.
In Section~5, I show that the geometric three dimensional action
can be interpreted as a first order formulation for
two dimensional chiral induced gravity,
and I comment on the generalization to \w s.
Although the nonlinearities present in \w s forbid a direct application of
the path integral method, the action obtained in Section~3 has natural
generalizations which are related to classical $W$-gravities.
There does not seem to be any three dimensional symmetry
in the $W$-case, but these actions do lead to some interesting results.
In particular, they define a set of $W$-curvatures, from which natural
geometric transformation rules for the $W$-gauge fields can be deduced.
Also, they provide a convenient framework for the explicit construction
of \w s, as discussed in [\walg].

\beginsection{2. \km\ algebras and gauged chiral \wzw s}
In this section I introduce a path integral method for constructing
topological actions
using an affine \km\ algebra as an example.
I recover the well known
relationship between \cs\ and chiral blocks of the \wzw. The method closely
parallels the construction of the geometric action on the coadjoint orbits
of a \km\ group [\AS,\Wieg], which has been described in detail in [\WK].

Let $G$ be a compact, connected, simply connected Lie group
with Lie algebra $\g$, and let \LG\ denote
the (universal) central extension of the loop group $LG$.
The Lie algebra $\lg$ of \LG\ is defined by the following commutator
$$[(u_1,m_1),(u_2,m_2)]=\bigl([u_1,u_2],{\textstyle{i\over2\pi}
  \tr\int dx\,u_2\d{x}u_1}\bigr).\eqno(2.1)$$
Here $(u(x),m)$, with $u:S^1\mapsto\g$ and $m\in\reals$,
denotes an element of $\lg$.

Consider a gauge potential $\Ah$ for $\lg$ defined on
a two dimensional surface $S$,
i.e. an $\lg$ valued one-form $\Ah=(A(x),a)$ on $S$.
Let $\ga$ be a contractible loop in $S$
and $\La$ an irreducible unitary representation
of \LG\ with highest weight $\la$ at level $k$,
and consider the Wilson loop
$\tr_{\La}\Pe{\oint_\ga\Ah}q^{\L_0}$, where $q\equiv\e{i\tau}$
and $\L_0\equiv i(\d{x}+\Ax)$ is a covariantized rotation generator.
$\Ax$ is a field on $S$ with values in the loop algebra
which under left translations $g\mapsto h^{-1}g$ transforms
in the coadjoint representation of the \km\ algebra
$$\Ax\mapsto h^{-1}\Ax h+h^{-1}\d{x}h\,.\eqno(2.2)$$
I will show that $\Ax$ can be related to
$\Ah$ by a constraint which is consistent with this transformation property.

In analogy with the \pin\ representation of a character
(see e.g. [\Stone]),
one can represent the \wl\  as a coherent state \pin\
over the \km\ group, whose elements are denoted by $\gh$
$$\tr_{\La}\Pe{\oint_\ga\Ah}q^{\L_0}=\int[d\gh]\e{S(\gh,\Ah)}
  =\int[d\gh]\e{\oint_\ga\bra\gh^{-1}d\gh+\gh^{-1}(\Ah+i\tau\L_0\,dt)\gh\ket}
  \,.\eqno(2.3)$$
Here, $\bra \hat u\ket$ denotes
the expectation value of $(u(x),m)\in\lg$ in the highest weight state
$$\bra\hat u\ket=\2dx\,\la u-km\,.\eqno(2.4)$$
Explicitly, the action in (2.3) is
$$\eqalign{S(\gh,\Ah)=&\two dx\,dt\bigl(\la g^{-1}(\dbar+\Abar)g+
     {\ts{ik\over2}}g^{-1}\d{x}(\dbar+2\Abar)g\bigr)\cr
			  &+\twelve_V(\mc)^3+\fourk dx\,dt\,
			  (-\tau\Ax^2)-k\oint_{\ga}a\,,\cr}\eqno(2.5)$$
where $\dbar\equiv\d{t}-\tau\d{x}$ and
$\Abar\equiv A_t-\tau\Ax$
is the chiral component of a \twod al gauge field minimally coupled to
the Noether current corresponding to the global invariance
under left translations of the ungauged action.
The two dimensional integral is over a torus $T=\ga\times S^1$
and $V=\Si\times S^1$
is the volume enclosed by $T$.

If $\Ah$ is taken to be a constant element of the \csa\ of $\lg$,
$\Ah=(H\,dt,0)$, with $H$ in the \csa\ of $\g$,
and $\Ax$ is set to 0,
the \wl\ reduces to
a \chr\ of $\lg$, and the action becomes that of a (twisted) chiral
\wzw. In that case, the \pin\ can be evaluated exactly and yields the
\wk\ \chr\ formula [\WK].

For general $\Ah$ and $\Ax$ it is not clear how to interpret
the last term in (2.5).
As I show now, for a subclass of these fields
it can be given an attractive interpretation
in terms of three dimensional field theory.

The field strength
$\Fh\equiv d\Ah+\half[\Ah,\Ah]$
can be written as
$$\Fh=(F(x),f)=(dA+A^2,da-\four dx\,A\d{x}A)\eqno(2.6)$$
and transforms in the adjoint representation
$$(F(x),f)\mapsto(h^{-1}Fh,f-\two dx\,\d{x}hh^{-1}F)\,.\eqno(2.7)$$
The transformation rule of the central part $f$ of the field strength
implies that it is consistent with
gauge invariance to impose the constraint
$$f=-\two dx\Ax F\,,\eqno(2.8)$$
where $\Ax$ is the field introduced above covariantizing $\d{x}$.
Combining (2.6) and (2.8) one has then
$$da=-\four dx(2\Ax F-A\d{x}A)\,.\eqno(2.9)$$
Given any $A(x)$ and $\Ax$, this constraint can be solved at least locally
for $a$.
On the other hand, it is clear that not every $\Ah$ admits a solution for
$\Ax$.
In the following, I will restrict to the subclass $\cal A$
of connections $\Ah$
for which there exists an $\Ax$ which solves (2.9).

The loop group part $A(x)$ of $\Ah$ defines two components of a
\3al gauge potential for the finite dimensional group $G$,
and $\Ax$ can be viewed as
a third component for this gauge field.
Thus, there is a correspondence between $\cal A$,
the subclass of those \twod al connections
for the \km\ group that admit a solution of (2.9),
and the \3al connections for the finite dimensional group.
The last term in (2.5),
$$\oint_\ga a=\int_\Si da=-{\ts{i\over4\pi}}\tr\intsx(2\Ax F-A\d{x}A)\,,
  \eqno(2.10)$$
can be expressed as a \3al action
for the \3al gauge field.
Notice that the arbitrariness in $\Ax$, given an $\Ah\in\cal A$,
corresponds to a gauge invariance of this action,
so that there is in fact a one-to-one correspondence
between gauge equivalence classes in $\cal A$ and gauge equivalence
classes of three dimensional connections.

Given the completely asymmetric way in which
$A_x$ and the remaining
two components have been introduced, one would not expect
any \3al symmetry between these components. Surprisingly,
the three components of $A$ enter (2.10) in a completely symmetric way.
In fact, (2.10) is, up to a boundary term,
exactly a \cs\ on the manifold $\Si\times S^1$.

Thus, for $\Ah\in\cal A$,
the action in the \pin\ expression  (2.3)
for the Wilson line
is a gauged chiral \wzw\ including a chiral Chern-Simons term
$$S(\gh,\Ah)=S_\la(g,\Abar)+S(A)\,,\eqno(2.11)$$
where
$$\eqalign{S_\la(g,\Abar)\equiv&\2dx\,dt\bigl(\la g^{-1}(\dbar+\Abar)g+
     {\ts{ik\over2}}g^{-1}\d{x}(\dbar+2\Abar)g\bigr)\cr
			      &\qquad+\twelve_V(\mc)^3\cr}\eqno(2.12)$$
and
$$S(A)\equiv\fourk dx\,dt\,\Abar\Ax-\fourk_V (AdA+{\ts{2\over3}}A^3)\,.
\eqno(2.13)$$

By construction, this action should be invariant under left translations
$g\mapsto h^{-1}g$, $A\mapsto h^{-1}Ah+h^{-1}dh$.
Indeed, both (2.12) and (2.13) are invariant up to a chiral \wzw\
(i.e.~an action of the form (2.12))
with $\la=0$, with opposite signs.
This is easily checked using the
chiral version of the \pw\ formula
$$S_\la(hg)=S_\la(g)+S_{g\la g^{-1}}(h)+\twok dx\,dt\,
  \d{x}g g^{-1}h^{-1}\dbar h\,,\eqno(2.14)$$
where
$$S_\la(g)\equiv\2dx\,dt\bigl(\la g^{-1}\dbar g+
     {\ts{ik\over2}}g^{-1}\d{x}\dbar g\bigr)+\twelve_V(\mc)^3\,.\eqno(2.15)$$

For comparison with the \vir\ case,
which will be discussed in the next section,
let me briefly summarize some points which show that
the action (2.11)
combines the main features of the relationship
between \cs\ and the \wzw\ in an interesting way.

The equations of motion  corresponding to (2.11) follow from
$$\eqalign{\de S(g,A)&=
  \2dx\,dt\bigl(-\Dbar(g\la g^{-1}+ik\d{x}gg^{-1})-ik\d{x}\Abar\bigr)
  \de gg^{-1}\cr
      &+\2dx\,dt(g\la g^{-1}+ik\d{x}gg^{-1}+ik\Ax)\de \Abar\cr
      &-\twok_VF\de A\,.\cr}\eqno(2.16)$$
Choosing a radial time coordinate,
perpendicular to the angular coordinate $t$
and the loop group coordinate $x$,
and performing canonical quantization on the torus $T$,
the physical states of the pure \cs\ are characterized by the \wi\ [\Witi]
$$F\vert_T=\Dbar \Ax-\d{x}\Abar=0\,,\eqno(2.17)$$
which is the projection of the \3al equation of motion on $T$.
Here, $\Ax$ and $\Abar$ are canonically conjugate variables
$$[\Abar,\Ax]={\ts{2\pi\over ik}}\,.\eqno(2.18)$$
It is then easily seen, using the equations of motion in (2.16)
and the gauge invariance of the measure,
that (2.17) is exactly the anomalous \wi\ satisfied by the functionals
$$\psi_\la(\Abar)\equiv\int [dg]\e{S_\la(g,\Abar)}\,.\eqno(2.19)$$
Indeed, it is well known that the $\psi_\la(\Abar)$ form a basis
for the Hilbert space $\cal H$ of physical states for \cs\ on
$T\times\reals$ [\BN].
As noted above, for $\Abar$ a constant element of the \csa\ of $\g$,
they reduce to \wk\ \chr s.

\beginsection{3. The \vir\ algebra}
The method presented in the previous section
can be applied to other centrally extended
Lie algebras. Here, I will consider the case of the \vir\ algebra.
\vir\ conformal blocks have been related to \3al \tft\ by H.~Verlinde,
who showed that in a specific, non gauge invariant polarization,
the physical state condition in $SL(2,\reals)$ \cs\ leads to the
\vir\ \wi\ [\V].
However, to carry out quantization
in such a non gauge invariant polarization seems problematic.

I will now construct an alternative topological action
which can be related to
\vir\ conformal blocks by following the procedure of the previous section.
The construction of this action  again
parallels that of the
geometric action on the coadjoint orbits [\AS,\Wieg]. I will denote elements
of the \vir\ algebra $\v$ by $(g(x),n)$ and elements of the dual algebra
$\v^*$ by $(b(x),c)$. Thus $g(x)\partial/\partial x$
is a vector field, while $b(x)\,dx^2$ is a quadratic differential
on the circle $S^1$.
The algebra is defined by the commutator
$$[(g_1(x),n_1),(g_2(x),n_2)]=(g_1g_2'-g_1'g_2,\ts{1\over48\pi}
  \int dx\,(g_1'''g_2-g_1g_2'''))\,,\eqno(3.1)$$
where the primes denote derivatives with respect to $x$, and the pairing
between $\v$ and $\v^*$ is
$$\bra(g(x),n),(b(x),c)\ket=\int dx\,g(x)b(x)+cn\,.\eqno(3.2)$$
The corresponding Lie group is the central extension of
the group of diffeomorphisms of  the
circle. It acts on $\v^*$ by the coadjoint action
$$\ad^*_G(b(x),c)=(b(G(x))G'^2-\ct S(G),c)\,,\eqno(3.3)$$
where $G:x\mapsto G(x)\in\diffs$ and $S(G)$
denotes the Schwarzian derivative of $G$.

A \chr\ of an irreducible unitary representation can be expressed as
a coherent state \pin\ by going through the same steps which lead to (2.3)
$$\tr q^{L_0}=\int[dG]\,\e{\oint d^{-1}\Om+\int dt\,\bra
  \ad_{G^{-1}}(i\tau L_0)\ket}\,,\eqno(3.4)$$
where $\Om$ is the symplectic form on the coadjoint orbit [\Witiii,\AS,\Wieg]
$$\Om(G)=\bra\bigl({dG\over G'}\bigr)^2\ket\,,\eqno(3.5)$$
and $\bra X\ket$ denotes the pairing of $X\in g$
with the coadjoint vector $(b_0,c)$.
For simplicity, I am choosing $b(x)=b_0$ to be a constant differential.
The explicit form of the action in (3.4) is
$$S=\int\xt(b_0G'\dbar G+\cf{\dbar G''\over G'})\,.\eqno(3.6)$$
For $\tau=0$, this action is equivalent to the one derived in [\AS,\Wieg].
As previously,
I introduce a gauge field
$\Ah=(\mu(x),a)$. Here $\mu(x)$, which is a gauge potential for $\diffs$,
transforms as a Beltrami differential
$$\mu\mapsto\mu\c HH'^{-1}-dHH'^{-1}\eqno(3.7)$$
under the substitution $G\mapsto G\c H$.
The corresponding field strength $\Fh\equiv d\Ah+\half[\Ah,\Ah]$
can be written as
$$\Fh=(F(x),f)=(d\mu+\mu\mu',da-{\ts{1\over48\pi}}\int dx\,\mu\mu''')
  \eqno(3.8)$$
and transforms in the adjoint \rep\
$$\eqalign{\Fh\mapsto\ad_H\Fh=&(F\c HH'^{-1},f-{\ts{1\over24\pi}}
  \int dy\,S(H^{-1})F)\cr
  =&(F\c HH'^{-1},f+{\ts{1\over24\pi}}\int dx\,S(H)F\c HH'^{-1})\,.\cr}
  \eqno(3.9)$$
The transformation law of $f$ implies that
it is consistent to impose a constraint
$$f={\ts{1\over24\pi}}\int dx\,TF\,,\eqno(3.10)$$
where $T$ transforms in the coadjoint representation
$$T\mapsto T\c HH'^2+S(H)\,.\eqno(3.11)$$
Eq.~(3.10) is the analogue of eq.~(2.8).
In the following, I restrict
to such $\Ah$ that (3.10) has a solution for $T$.

The \pin\ expression for a \wl\ for $\Ah$ is analogous to (3.4)
$$\tr P\e{\oint\Ah}=\int[dG]\,\e{\oint (d^{-1}\Om+
  \bra\ad_{G^{-1}}\Ah\ket)}\,.\eqno(3.12)$$
The second term in the action
$$\oint\bra\ad_{G^{-1}}\Ah\ket=\int\xt\,\mu(b_0G'^2-
  \ct S(G))+c\oint a\eqno(3.13)$$
contains, besides a coupling to the Noether current,
again a term involving the central part of
the gauge potential. Using the same trick as in the \km\ case, I write
this as a \3al action
$$c\oint a=c\int_\Si da=\cf\intsx(2TF+\mu\mu''')\,.\eqno(3.14)$$
The action in the \pin\ (3.12) can now be written explicitly as
the sum of a \twod al and a \3al part
$$S(G,\hat\mu)=S_2(G,\mu_t)+S_3(\mu,T)\,,\eqno(3.15)$$
with
$$S_2(G,\mu_t)=\int\xt(b_0G'(\t{G}+\mu_t G')+\cf({\t{G}''\over G'}
  +2\mu_t S(G))\eqno(3.16)$$
and
$$S_3(\mu,T)=\cf\intsx(2TF+\mu\mu''')\,.\eqno(3.17)$$
\medskip

Again, it can be shown that the total action is gauge invariant,
although this is less straightforward than in the \km\ case.
The analogue of the \pw\ formula is
\smallskip
$$\eqalign{\int\xt{\t{(G\c H)}''\over (G\c H)'}=
	  &\int\xt{\t{G}''\over G'}\c HH'+\int\xt{\t{H}''\over H'}\cr
	  &\qquad-2\int\xt S(G)\c H\t{H}H'\,,\cr}\eqno(3.18)$$
from which it follows that
$$S_2(G\c H,\mu_t\c HH'^{-1}-\t{H}H'^{-1})=S_2(G,\mu_t)
       +S_2(H,\mu_t\c HH'^{-1}-\t{H}H'^{-1})\vert_{b_0=0}\,.\eqno(3.19)$$
\smallskip\noindent
As in the \km\ case, the \3al action is invariant up to a boundary term.
This boundary term is precisely the two dimensional action
without the
$b_0$ term
$$\eqalign{&S_3(\mu\c HH'^{-1}-d{H}H'^{-1},T\c HH'^2+S(H))=\cr
	   &\qquad S_3(\mu,T)-S_2(H,\mu_t\c HH'^{-1}-
	    \t{H}H'^{-1})\vert_{b_0=0}\,,\cr}\eqno(3.20)$$
and cancels the second term in (3.19).
\smallskip\noindent

The equations of motion corresponding to (3.15) follow from
$$\eqalign{\de S&=\int\xt(-\nabla_t(b_0G'^2-\ct S(G))+\ct\mu_t''')
   {{\de G\over G'}}\cr
		&\qquad+\int\xt(b_0G'^2-\ct S(G)+\ct T)\de \mu_t\cr
		&\qquad+\ct\intsx(\nabla T+\mu''')\de \mu\,.\cr}\eqno(3.21)$$
Here, $\nabla\phi$ denotes the covariant derivative
$$\nabla\phi\equiv d\phi +\mu\phi'+s\mu'\phi\eqno(3.22)$$
of a field with spin $s$, i.e., the homogeneous part
of whose transformation rule is
$\phi\mapsto\phi\c H(H')^s$.
Notice that (3.21) is completely analogous to the
corresponding equation in the \km\ case (2.16).

Choosing the radial coordinate of the disk $\Si$ as time coordinate,
and performing canonical quantization, one finds that $T$ and $\mu_t$ are
canonically conjugate
$$[\mu_t,T]={\ts{24\pi\over c}}\,.\eqno(3.23)$$
The physical state condition following from (3.21)
$$(\nabla T+\mu''')\vert_\Si=0\eqno(3.24)$$
is the \vir\ \wi\ [\V].
This is the defining equation for conformal blocks of
the partition function for
conformal field theories
whose maximal chiral algebra is the \vir\ algebra.
The partition function of the two dimensional action
$$\psi(\mu_t)\equiv\int[dG]\,\e{S_2(G,\mu_t)}\eqno(3.25)$$
is easily seen, using (3.21), to solve the \wi\ (3.24).
Thus, it defines a physical state of the \3al theory.

\beginsection{4. Covariant form of the \3al action}
In the previous section,
I obtained a \3al \tft\ associated to the \vir\ algebra
coupled to the chiral part
of a two dimensional \cft\
on the boundary.
This two dimensional theory defines states
in the Hilbert space associated to
canonical quantization of the \3al theory.

At this point, the \3al action (3.17) looks rather asymmetric
between the Beltrami differential $\mu$ and the field $T$.
I will show now that there is in fact a remarkable
symmetry between the two, and that recognizing this symmetry
enables one to rewrite the
action $S_3$ in a manifestly generally covariant form.

I restricted to those \vir\ connections which can be parametrized as
$$\Ah\equiv(\mu,d^{-1}(\cf\int dx\,(2TF+\mu\mu''')))\,.\eqno(4.1)$$
The only relevant property of $T$ in this expression is its transformation
rule (3.11), which says that $T$ transforms as a stress tensor.
It is well known that one can parametrize $T$
in terms of a so called
\ac\ $\la$ (see e.g.~[\IZ])
$$T\equiv-(\la'+\half\la^2)\,,\eqno(4.2)$$
where $\la$ transforms in the following way
$$\la\mapsto\la\c HH'-{H''\over H'}\,.\eqno(4.3)$$

The \ac\ $\la$
can be used to define covariant derivatives with respect to $x$
on objects  that transform as tensors under $\diffs$
(remember that $\mu$ defines covariant {\it exterior}
derivatives, cf.~(3.22)).
Indeed, it is easily checked that for $\phi$ a tensor of spin $s$
$$D\phi\equiv\phi'+s\la\phi\eqno(4.4)$$
is a tensor of spin $s+1$.

{}From $\la$ and $\mu$, one can construct a \3al gauge field
$$A\equiv D\mu+\la\,dx\,.\eqno(4.5)$$
Using (3.7) and (4.3), it is straightforward
to verify that $A$ transforms as
$$A\mapsto A-d\log{H'}\,,\eqno(4.6)$$
where $d$ is now the exterior derivative in \3s,
$d\equiv d\vert_{\Si}+dx\,\d{x}$.
Moreover, up to a boundary term, the action (3.14) can be rewritten as an
abelian Chern-Simons action
$$S_3(\mu,T)=\int_V AdA-\int\xt\mu_t'\la\,.\eqno(4.7)$$
Hence, there is a very remarkable symmetry in \3s between the
Beltrami differential $\mu$
and the \ac\ $\la$.

Obviously, the \wl s
$$\tr_s P\e{\oint A}=P\e{s\oint A}\eqno(4.8)$$
are gauge invariant observables of the \3al theory.
Canonical quantization on a space $\Si$,
pierced by a collection of such \wl s corresponding to spins $s_i$,
leads to the following \wi\
$$\nabla_t T+\mu_t'''
  +\tc\sum_i(s_i\de'(z-z_i)+\de(z-z_i){\partial/\d{x_i}})=0\,,\eqno(4.9)$$
where $z=(x,t)$ and $z_i=(x_i,t_i)$ denote the positions of the punctures.
This is indeed the correct \wi\ defining the \vir\ conformal blocks [\V].
The states satisfying this \wi\ can be expressed as
$$\psi(\mu_t,z_i)\equiv\int[dG]\prod_i(G'(z_i))^{s_i}\e{S_2(G,\mu_t)}
  \,.\eqno(4.10)$$
One would expect, by analogy with the \wzw, that
these \wl s represent the Verlinde operators, and can be used to
construct knot invariants.

\beginsection{5. Classical W-gravity}
The \eom\ (3.21) of the \vir\ action are solved
by the pure gauge configurations
$\mu=-{dH\over H'}$, $T=S(H)$.
Substituting these expressions in the \3al action (3.17)
and using the transformation property (3.20),
one sees that the action reduces to a pure boundary term
which is exactly Polyakov's induced gravity in the chiral gauge [\P].
In this sense, the \3al action  can be interpreted as
a first order formulation of
Polyakov's gravity.
This can also be seen more directly as follows. Upon applying
the \3al \eom, the variation of the \vir\ action is a boundary term
$$\de S=-\ct\intxt T\de\mu=\ct\intxt\nabla^{-1}\mu'''\de\mu\,.
  \eqno(5.1)$$
This is the anomaly defining the induced action.

The induced action for $W$-gravity theories has a rather complicated
structure, and in particular
includes nonlocalities in the quantum case [\SvN]. The \3al
first order formulation  of induced gravity, as discussed above,
can be extended straightforwardly
to first order formulations of classical $W$- and super $W$-gravities.
A detailed construction of such extensions is given in [\walg].
In the following, I will discuss the example of $W_3$-gravity.
The action in this case is [\walg]
$$S=\intsx(2TF+2W\nabla\nu+\mu\mu'''+\nu\nu^{(5)}+10\nu\nu'''T-6\nu\nu'T''
    +16\nu\nu'T^2)\,.\eqno(5.2)$$
Here, $W$ is the spin 3 field and $\nu$ is a spin $-2$ one-form,
the gauge field of the $W$-symmetry.
The \eom\ are the classical $W_3$ Ward identities (see e.g.~[\SvN])
$$\nabla T+\mu'''+3\nu'W+2\nu W'=0\,,\eqno(5.3)$$
$$\nabla W+D^5\nu=0\,,\eqno(5.4)$$
where $D^5$ is the second \gd\ operator, and the Maurer-Cartan equations
$$d\mu+\mu\mu'+2\nu\nu'''-3\nu'\nu''+16\nu\nu'T=0\,,\eqno(5.5)$$
$$\nabla\nu=0\,.\eqno(5.6)$$

Eqs.~(5.5,6) define an algebra with field dependent
structure constants. Together with eqs.~(5.3,4)
they form an integrable system, i.e.,
the integrability condition
$d^2=0$ is identically satisfied.

In the standard way one derives the gauge
transformation laws for the gauge potentials $\mu$ and $\nu$
$$\de\mu=\nabla\ep+2\nu\eta'''-2\nu'''\eta-3\nu'\eta''
	 +3\nu''\eta'+16\nu T\eta'-16\nu'T\eta\,,\eqno(5.7)$$
$$\de\nu=\nabla\eta+2\nu\ep'-\nu'\ep\,.\eqno(5.8)$$
Similarly, the Ward identities (5.3,4) determine, by contraction
with the gauge parameters,
the standard transformation laws of $T$ and $W$
$$\de T=-2\ep'T-\ep T'-\ep'''-3\eta'W-2\eta W'\,,\eqno(5.9)$$
$$\de W=-3\ep'W-\ep W'-D^5\eta\,.\eqno(5.10)$$
The variation of the action under these gauge transformations is
$$\de S=-2\intxt(\ep\mu'''+\eta(\nu^{(5)}-16\nu'T^2-16\nu TT'))
  \,.\eqno(5.11)$$

The transformation laws (5.7--10)
lead to the following transformation laws for the curvatures (5.3--6)
$$\eqalign{\de\F&=-\ep\F'+\ep'\F-2\eta\G'''+2\eta'''\G
		  +3\eta'\G''-3\eta''\G'\cr
		&\quad-16\eta\G'T+16\eta'\G T
		  +16\eta\nu'\T-16\eta'\nu\T\,,\cr}\eqno(5.12)$$
$$\de\G=2\ep'\G-\ep\G'-2\eta\F'+\eta'\F\,,\eqno(5.13)$$
$$\de\T=-2\ep'\T-\ep(\T)'-3\eta'\W-2\eta(\W)'\,,\eqno(5.14)$$
$$\de\W=-3\ep'\W-\ep\W'\,,\eqno(5.15)$$
where $\T$, $\W$, $\F$ and $\G$ are the left hand sides of (5.3--6),
respectively.
As in an ordinary Lie algebra, the curvatures transform homogeneously,
so that the \eom, which state the vanishing of these curvatures,
are gauge covariant. Hence, eqs. (5.9,10) are still valid in
second order formalism, in which $T$ and $W$ become functions of $\mu$
and $\nu$ so that their transformation laws are no longer independent.
The variation (5.11)
is thus in fact the variation of the induced gravity action.
All this is in agreement with the classical limit of
the corresponding results obtained in [\SvN] for quantum induced gravity.

\beginsection{6. Conclusion}
I have presented in this \note\ \3al actions related to both
linear and nonlinear chiral algebras
of importance in conformal field theory, as well as a general procedure
to construct such actions in the linear case.
A simpler procedure to construct
these actions  which generalizes
to nonlinear algebras is described in [\walg].
The \eom\ corresponding to these actions include the
Ward identities of the algebra, as well as Maurer-Cartan equations
for the gauge potentials.
The action associated to the \vir\ algebra is a first
order formulation of \twod al induced gravity in the chiral gauge,
while the action associated to the $W_3$-algebra is a first order formulation
of the classical limit of $W_3$-gravity.
The \vir\ and \km\ actions have a surprising \3al symmetry,
the latter being the action for \cs\ for the corresponding
finite dimensional group. The \vir\ action can be written
in the form of an abelian \cs,
and admits the definition of \wl s. These \wl s create insertions
in the conformal blocks, and might be of interest in knot theory.

\beginsection{Acknowledgements:}I would like to thank K.~Pilch for
suggesting many improvements to the manuscript and for
illuminating discussions. This research was partially
supported by the Department of Energy Contract \#DE--FG03--91ER40168.


\def\eef #1 {\llap{#1\enspace}\ignorespaces}
\def\ref{\beginsection{References}\frenchspacing
    \parindent=0pt\leftskip=1truecm\parskip=8pt plus 3pt
    \everypar={\eef}}
    \def\npb#1{Nucl. Phys. {\bf B#1}}
    \def\cmp#1{Commun. Math. Phys. {\bf#1}}

    \def\plb#1{Phys. Lett. {\bf #1B}}
    \def\ijmp#1{Int. J. Mod. Phys. {\bf A#1}}
    \def\jgp#1{J. Geom. Phys. {\bf#1}}
    
    \def\mpl#1{Mod. Phys. Lett. {\bf A#1}}

\def\hb{\hfil\break}

\ref

[1] E. Witten, \cmp{121} (1989) 351

[2] E. Witten, \npb{322} (1989) 629; \npb{330} (1990) 285

[3] M. Bos and V.P. Nair, \plb{223} (1989) 61; \ijmp{5} (1990) 959\hb
    S. Elitzur, G. Moore, A. Schwimmer and N. Seiberg, \npb{326} (1989) 108

[4] E. Verlinde, \npb{300} (1988) 360

[5] A. Alekseev, L. Faddeev and S. Shatashvili, \jgp{1} (1989) 3

[6] M. Stone, \npb{314} (1989) 557

[7] A. Alekseev and S. Shatashvili, \npb{323} (1989) 719

[8] P.B. Wiegmann, \npb{323} (1989) 311

[9] R.E.C. Perret, \npb{356} (1991) 229

[10] R.E.C. Perret, USC--92/016 ({\tt hepth/9208068})

[11] H. Verlinde, \npb{337} (1990) 652

[12] E. Witten, \cmp{114} (1988) 1

[13] M. Bauer, P. Di Francesco, C. Itzykson and J.-B. Zuber,
     \npb{362} (1991) 515

[14] A.M. Polyakov, \mpl{2} (1987) 893

[15] H. Ooguri, K. Schoutens, A. Sevrin and P. van Nieuwenhuizen,
     \cmp{146} (1992) 616

\vfil\eject\end